\documentclass[journal=jacsat,manuscript=article]{achemso}
\setkeys{acs}{keywords = true}

\usepackage{chemformula} % Formula subscripts using \ch{}
\usepackage[T1]{fontenc} % Use modern font encodings

\author{Xin Zhou}
\affiliation{Univ. Lille, CNRS, Centrale Lille, Univ. Polytechnique Hauts-de-France, UMR 8520 - IEMN, F-59000 Lille, France }
\email{xin.zhou@iemn.fr}
\author{Srisaran Venkatachalam}
\affiliation{Univ. Lille, CNRS, Centrale Lille, Univ. Polytechnique Hauts-de-France, UMR 8520 - IEMN, F-59000 Lille, France }
\author{Ronghua Zhou}
\affiliation{Univ. Lille, CNRS, Centrale Lille, Univ. Polytechnique Hauts-de-France, UMR 8520 - IEMN, F-59000 Lille, France }
\author{Hao Xu}
\affiliation{Univ. Lille, CNRS, Centrale Lille, Univ. Polytechnique Hauts-de-France, UMR 8520 - IEMN, F-59000 Lille, France }
\author{Alok Pokharel}
\affiliation{Univ. Lille, CNRS, Centrale Lille, Univ. Polytechnique Hauts-de-France, UMR 8520 - IEMN, F-59000 Lille, France }
\author{Andrew Fefferman}
\affiliation{Univ. Grenoble Alpes, Institut NEEL - CNRS UPR2940, 25 rue des Martyrs, BP 166, 38042 Grenoble Cedex 9, France}
\author{Mohammed Zaknoune}
\affiliation{Univ. Lille, CNRS, Centrale Lille, Univ. Polytechnique Hauts-de-France, UMR 8520 - IEMN, F-59000 Lille, France }
\author{Eddy Collin}
\affiliation{Univ. Grenoble Alpes, Institut NEEL - CNRS UPR2940, 25 rue des Martyrs, BP 166, 38042 Grenoble Cedex 9, France}

\title[An \textsf{achemso} demo]
  {High-Q silicon nitride drum resonators strongly coupled to gates}


\begin{document}


\begin{abstract}
Silicon nitride (SiN) mechanical resonators with high quality mechanical properties are attractive for fundamental research and applications. However, it is challenging to maintain these mechanical properties while achieving strong coupling to an electrical circuit for efficient on-chip integration. Here, we present a SiN drum resonator covered with an aluminum thin film, enabling large capacitive coupling to a suspended top-gate. Implementing the full electrical measurement scheme, we demonstrate a high quality factor $\sim 10^{4}$ (comparable to that of bare drums at room temperature) and present our ability to detect $\sim 10$ mechanical modes at low temperature. The drum resonator is also coupled to a microwave cavity, so that we can perform optomechanical sideband pumping with a fairly good coupling strength G and demonstrate mechanical parametric amplification. This SiN drum resonator design provides efficient electrical integration and exhibits promising features for exploring mode coupling and signal processing.

\end{abstract}

\section{Introduction}
%This is a paragraph of text to fill the introduction of the
%demonstration file.  The demonstration file attempts to show the
%modifications of the standard \LaTeX\ macros that are implemented by
%the \textsf{achemso} class.  These are mainly concerned with content,
%as opposed to appearance.
%Nano Letters is a communications journal for publishing results that need to be distributed rapidly. Therefore, the Letter should contain no more than 5 figures and contain no more than 3000 words. (Note: this does not include references, abstract, or captions) The abstract is limited to 150 words. Manuscripts that are not within these guidelines will be returned for resubmission before the Letter will be considered by the Editors. Nano Letters is pleased to publish papers without page or color charges to authors.

Silicon nitride (SiN) strings and membranes, fabricated from pre-stressed thin films, have emerged as promising devices \citep{zwickl2008high, verbridge2006high}. They have nanogram (ng) effective mass $m_{eff}$ and easily achieve MHz range resonance frequency with high quality factor $Q$ \cite{adiga2011modal}, as the pre-stress is well known for diluting the dissipation \cite{verbridge2006high, unterreithmeier2010damping}. SiN based micro- and nano-electromechanical systems (MEMS and NEMS) allow electrical signals to couple with a mechanical degree of freedom and give access to electrical integration on-chip. They are of interest for both fundamental science and applications, from room temperature (RT) to mK temperature. Because of those unique properties mentioned above, SiN nano-electromechanical resonators have been exploited through their coupling to a microwave cavity, implementing an electric analog of optomechanics \citep{aspelmeyer2014cavity, zhou2021electric}. Based on this platform, a mechanical resonator has been cooled down to microkelvin temperature, been exploited for on-chip thermometry, and been built as a highly sensitive detection scheme at RT \cite{yuan2015large,WeigCavityRT,zhou2019chip}.  

However, the insulating feature greatly limits implementations of SiN mechanical resonators in electrical systems, because a purely dielectric actuation and detection scheme is particularly weak. Up to now, composite SiN doubly-clamped beams are one of the simplest and widely used device structures, in which the suspended beam is covered with a thin metal layer to generate the capacitive coupling with a side-gate. Ameliorations of coupling capacitance, for a beam structure, are normally achieved by increasing its length through lowering the resonance frequency ($\Omega_m$) or using a demanding technique to reduce the vacuum gap  \cite{massel2011microwave}. Non-metallized SiN beams, driven by dielectric force, provide an alternative electrical integration scheme with a high Q feature $\sim 10^5$ at RT \cite{unterreithmeier2009universal}. Unfortunately, its typical coupling factor is quite low in microwave optomechanical platforms $\sim$ 70 Hz/nm  \cite{WeigCavityRT}. Therefore this MEMS/NEMS conception based on a doubly clamped beam structure leaves limited space for making trade-offs among the coupling strength, $\Omega_m$ and $Q$. For this reason, SiN membranes with the specificity of a high surface-to-volume ratio motivate researchers to explore electrical integration, because they can have strong coupling with the surrounding electrical circuits. A SiN-Graphene electromechanical resonator has been demonstrated with an electrical readout through a back-gate structure \cite{lee2013graphene}. In very recent years, flip-chip techniques are used to achieve electrical integration through creating a capacitive coupling between a SiN membrane coated with an aluminum (Al) thin layer and a 3-dimensional microwave cavity \cite{yuan2015large}. Although these previous works pave the way toward the electrical integration, it is still a challenge for these SiN membrane devices to serve in today's microelectronics and quantum engineering. In these fields, it is essential to build scalable architectures, to achieve multiple device integration, and to have well controlled capacitive coupling. Thus, there is a strong motivation to explore new types of membrane resonators, with large electrical coupling effects and good mechanical properties at RT and low temperature.

In this work, we report on the electrical integration of a SiN drum resonator with microwave reflectometry and a microwave cavity forming an optomechanical platform. We present a scalable SiN drum resonator capacitively coupled to its suspended top-gate, which has been achieved through a standard top-down nanofabrication process. Both low and high stress drums were studied. This ultra-clean process yields quality factors $Q\sim 10^4$ at RT, which is comparable with that of bare SiN drums \cite{adiga2012approaching}. By using microwave reflectometry, we are able to characterize the basic mechanical properties, investigate Duffing nonlinearity, and detect 10 mechanical modes. By means of the optomechanical platform, we clearly demonstrate optical damping through sideband pumping of  the cavity, giving coupling strength values in accordance with our nanofabrication parameters. These studies indicate that this new type of SiN drum resonators can be an excellent device for studying mechanical mode coupling, exploring signal processing and quantum engineering.

\section{DEVICE FABRICATION AND MEASUREMENT SETUP}
\begin{figure}
  \includegraphics[width=0.95\textwidth]{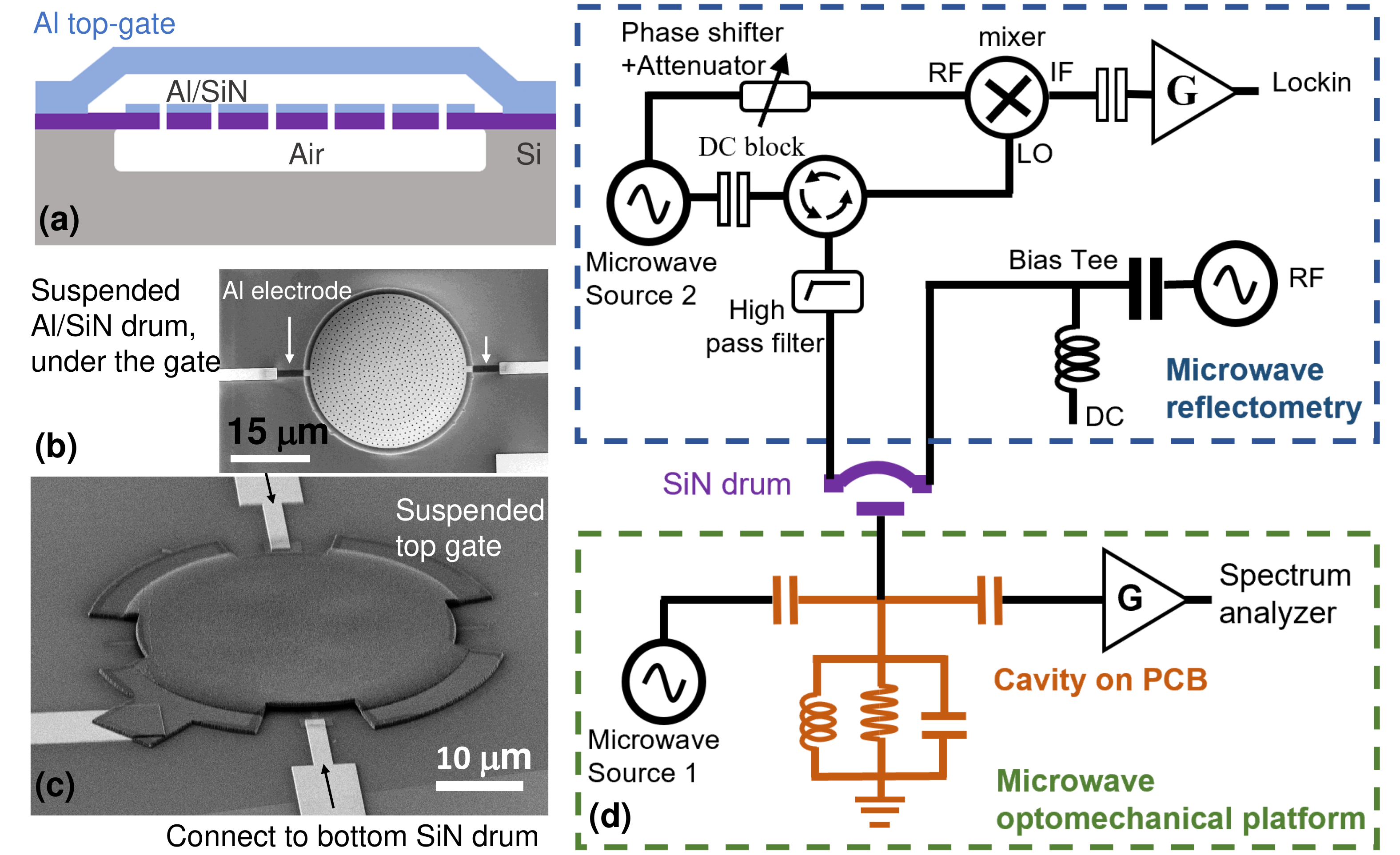}
  \caption{(a) schematic diagram of lateral device structure, (b) SEM image of the suspended SiN drum covered with 25 nm Al thin film. In order to minimize the contribution of the metal to the damping, it is absent in most of the clamping region \cite{yu2012control}. This Al thin film is connected with the outside electrodes through two rectangle Al electrodes. (c) SEM image of the final device structure, in which Al/SiN capacitively couples with a suspended top-gate, and (d) schematic diagram of measurement setup, in which the microwave cavity on the PCB part is marked in caramel color. The microwave cavity makes capacitive coupling with SiN drum (purple color) via bonding wires connecting to its suspended top-gate. The SiN drum, covered with Al thin film, connects to two microstrip transmission lines through bonding wires. One is for driving the mechanical resonator and the other one is for detecting mechanical motion through microwave reflection scheme \citep{legrand2013detecting}. More details are shown in the supporting information (SI:Nanofabrication, The setup and modeling of microwave reflectometry).}
  \label{sch:setup}
\end{figure}

  The device fabrication process starts with a silicon substrate ($\sim$ 10k $\Omega$-cm) covered with $\sim$ 100 nm stoichiometric SiN thin film. Inspired by previous work of SiN drum \cite{adiga2012approaching}, we define diameter of the drum by using electron beam (EB) lithography to pattern circularly symmetric holes with 200 nm in diameter and 1.5 $\mu$m in spacing distance on the  CSAR 62 resist coated on a SiN/silicon wafer \citep{thoms2014investigation}. We take this resist as an etching mask and the drum is released from the Si substrate by reactive ion etching of the SiN layer followed by a selective XeF$_2$ silicon etching through these holes. An Al thin film, around 25 nm in thickness, is deposited on the top of the suspended SiN drum, as shown in Fig.\ref{sch:setup}.(b). The drum's suspended top-gate is fabricated by using PMMA (polymethyl methacrylate) resist, $\sim 500$ $\pm$ 50 nm, as a top-gate support through soft-bake and reflowed process \citep{abuwasib2013fabrication}. Then, the top-gate design is patterned on the second layer of MMA (methyl methacrylate) resist, deposited on the top of the PMMA layer. The fabrication process is finished by deposition of $\sim$ 600 nm Al thin film followed with lift-off process. This lift-off process also removes the gate support resist, forming a suspended top-gate structure. Figure \ref{sch:setup}.(c) shows a SEM (scanning electron microscope) image of the final device, in which SiN drum in the bottom is connected with two electrodes and capacitively coupled with its suspended Al gate on the top. In this work, we fabricate SiN drum with a low ($\sim$ 0.6 GPa) and a high ($\sim$ 1.0 GPa) tensile stress SiN thin film, respectively. 

The experimental setup is schematically depicted in Fig. \ref{sch:setup} (d). The SiN drum is electrically integrated with a printed circuit board (PCB), which is designed to be compatible with both microwave optomechanical platform and reflectometry measurement scheme, as illustrated in Fig. \ref{sch:setup} (d).  We are using a quarter wavelength ($\lambda$/4) resonator, shorted microstrip transmission line, to build a microwave cavity on PCB \cite{WeigCavityRT}. It has a resonance frequency $\omega_c \sim$ 5.2 GHz with a quality factor $Q_c \sim$ 85 at RT, which is measured by standard transmission scheme.  All the measurements are performed under vacuum ($\sim 10^{-6}$ mbar) to minimize air damping.   
\subsection{RESULTS AND DISCUSSIONS}
\begin{figure}
  \includegraphics[width=0.98\textwidth]{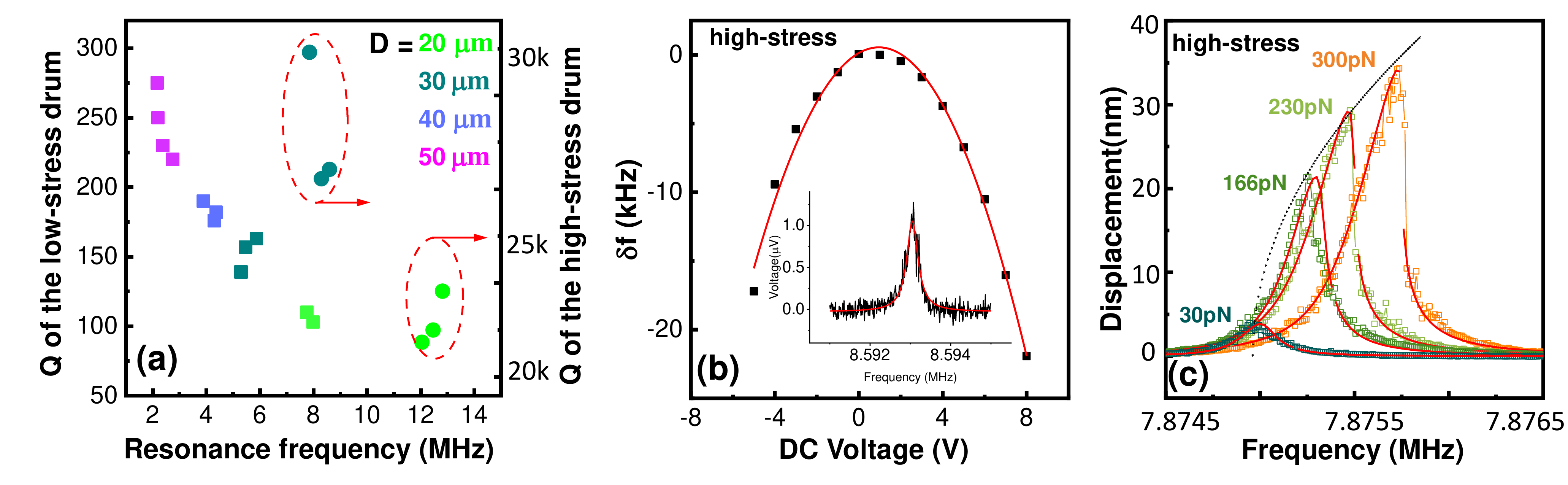}
  \caption{Fundamental modes of SiN drums, measured at RT. (a) Quality factors of low- (square) and high-stress (circle) drum resonators with diameters from $D$=20 $\mu$m to 50 $\mu$m, corresponding to different colors. The mechanical responses are measured at $V_{dc}$=0, $V_{ac}$=15 mV and 0.5 mV for low- and high-stress drums. (b) Resonance frequency of a high-stress SiN drum ($D$=30 $\mu$m) under modulation of the $V_{dc}$ when the device is driven by an $ac$ signal with an amplitude of $V_{ac}$=0.5 mV.  The red line is a quadratic fit and the inset figure is a resonance response corresponding to $V_{dc}$=0. (c) Nonlinear response of a high-stress SiN drum. Both black and red lines are fitting curves (see the text and SI:Electrostatic force on frequency modulation, Nonlinear fit Duffing parameter)}
  \label{sch:LHres}
\end{figure}
Our drum resonators have diameters of order 10 to 100  $\mu$m and thicknesses $\sim$100 nm. Therefore, they can be accurately modeled as circular membranes. The eigenfrequency of such a membrane is strongly dependent on its diameter $D$ and tensile stress $\sigma$, and is given by 
\begin{equation}
  \Omega_{nm}=2 \zeta_{nm} c_R/D. \label{eqn:eigenfreq}
\end{equation}
Here, the eigenfrequency $\Omega_{n,m}$ is designated by ($n$, $m$), where $n$ index represents the azimuthal mode number and the $m$ index denotes the radial mode number. The $c_R=\sqrt{\sigma/\rho}$ is the phase velocity, $\rho$ is the effective mass density, and $\zeta_{nm}$ is the $m$th zero of Bessel function \cite{graffwave}. In this experiment, mechanical vibrations of the suspended Al top-gate have been observed in few devices, exhibiting resonant frequencies and quality factors much lower than their coupled SiN drums, due to the low residual stress in the Al films \citep{guisbiers2006residual, cattiaux2020geometrical} and the large clamping losses. We, therefore, focus on investigations of the SiN drum resonators through electrically driving them around their resonance frequencies, avoiding unwanted effects from their suspended gates. We first characterize the fundamental mode $\Omega_{01}$ of the resonators at RT. The measurement is performed by applying a $dc$ signal $V_{dc}$ combined with an ac signal $V_{ac}$ to drive the resonator and by utilizing the microwave reflectometry to detect their mechanical responses, as shown in Fig.\ref{sch:setup} (d). We measure several resonators with different $D$ varying from 20 $\mu m$ to 50 $\mu m$ and observe that both $\Omega_{01}$ and $Q$ exhibit clearly a size dependence, as expected, for the low-stress and the high stress drums, as shown in Fig. \ref{sch:LHres} (a). By using Eq.\ref{eqn:eigenfreq}, the average phase velocities are evaluated to be $\approx$ 212 m/s and $\approx$ 313 m/s for the low and the high stress SiN drum, which could be attributed to a reduction of tensile stress generated by our specific drum release geometry \citep{wilson2011high}. Besides, the Q value decreases as the diameter of the drum decreases, because clamping losses dominate mechanical dissipation. The average $Q$ of the high stress devices reaches $\sim 10^4$, two orders of magnitude higher than that of low stress resonators. It is well known in studies of SiN beam and membrane resonators that high tensile stress can result in higher $Q$, which has been attributed to the enhancement of the stored energy from the stress \cite{verbridge2006high, yu2012control}. As discussed above, the quality factor of a vibrating membrane depends on device dimension and tensile stress of the material. We therefore choose results reported in the literature from a SiN  membrane with similar dimensions and tensile stress to make a reasonable comparison. In our work, the quality factor of a high-stress drum covered with 25 nm Al thin film ($D$ = 20 $\mu$m) reaches the value of $\sim$2.4 $\times$ 10$^4$ at $\textit{RT}$, which arrives at the value of a bare SiN drum detected by optical interferometry  \cite{adiga2012approaching}. Our ultra-clean nanofabrication process enables SiN drums to have capacitive coupling without adding additional dissipation to the device.

Figure \ref{sch:LHres}.(b) shows typical resonance responses of a SiN drum resonator as a function of the voltage $V_{dc}$ applied to the drum. The quantity $\delta f$ is the shift of the resonance frequency relative to the maximum value, where 2$\pi \delta f$ = $\Omega_{01}$ - $\Omega_{max}$. It follows a quadratic behavior stemming from the electrostatic force $F(x)=\partial (C_g(x)V^2/2)/\partial x$, where $V$ = $V_{dc}+V_{ac}$ and $x$ is the displacement of the membrane. The bias voltage shifts $\Omega_{01}$, by changing the intrinsic spring constant of the drum resonator $k$, described by $\Omega_m \approx \Omega_{01} (1+\delta k/(2k))$. The $\delta k \approx  - \partial F/\partial x$ is the small variation of the $k$. 
If we consider the case $V_{dc} \gg \vert V_{ac} \vert $,  the shifts of $\Omega_{01}$ are mainly dominated by the $V^2_{dc}$ term in the expression for $F$. The 2$V_{dc}V_{ac}$ and $V^2_{ac}$ terms can be exploited to drive the resonator with a force oscillating at frequency $\sim \Omega_m$ or to perform parametric amplification by modulating $k$ at frequency $\sim 2 \Omega_m$ (see more details in SI). The gate coupling scheme gives thus access to a diversity of operations of mechanical resonators. The measurement results are fit based on these expressions and we obtained the average intrinsic spring constant $k \approx$17 N/m and $\approx$ 90 N/m for the low- and high-stress SiN drum with $D = 30$ $\mu m$, respectively.

Nonlinearity is also one of the important features of a mechanical resonator in applications of signal processing and sensing. Therefore, we also drive the fundamental mode (0,1) of a drum resonator ($D$ = 30 $\mu$m) to the nonlinear region through upward frequency sweeping, as shown in Figure \ref{sch:LHres}(c). Here, the displacement of the drum is recalculated from the amplitude of electrical signals in microwave reflectometry (see details in SI) and the driving  force was created by varying $V_{ac}$ from 0.5 mV to 4 mV and keeping $V_{dc}$ = 2 V $\gg \vert V_{ac} \vert $. The nonlinear mechanics of a mode can be modeled by a well-known motion equation
\begin{equation}
  \ddot{x}+\gamma_{m}\dot{x}+\frac{k}{m_{eff}}x+\alpha x^3=\frac{F cos(\omega t)}{m_{eff}}, \label{eqn:nonlinear}
\end{equation}
where $m_{eff}$ is the effective mass of the mode, $x$ is mechanical displacement and $\alpha$ is the Duffing parameter. As mentioned above, the $F$ is the driving force. For the first mode (0,1), $m_{eff}$ is calculated by taking the proper mode shape factor 0.269 to multiply the drum's mass \citep{hauer2013general}. The measurement results are fit by utilizing secular perturbation theory to solve the Eq.\ref{eqn:nonlinear} and treat the drive force, $Q$ and $\alpha$ as parameters to be fit. The best fitting results are attained within $\sim$ 5$\%$ error bar, which give a reasonable Duffing parameter (normalized to the mode mass, $\sim$ 3.4$\times 10^{-14}$ kg) $\alpha \approx$ 3.2 $\times 10^{26} m^{-2}s^{-2}$
. This value is however about 10 times smaller than the simple theoretical estimate  \citep{cattiaux2020geometrical}, which we could attribute to the tiny holes structure on the membrane which certainly disturb the biaxial stress pattern. 
\begin{figure}
  \includegraphics[width=0.5\textwidth]{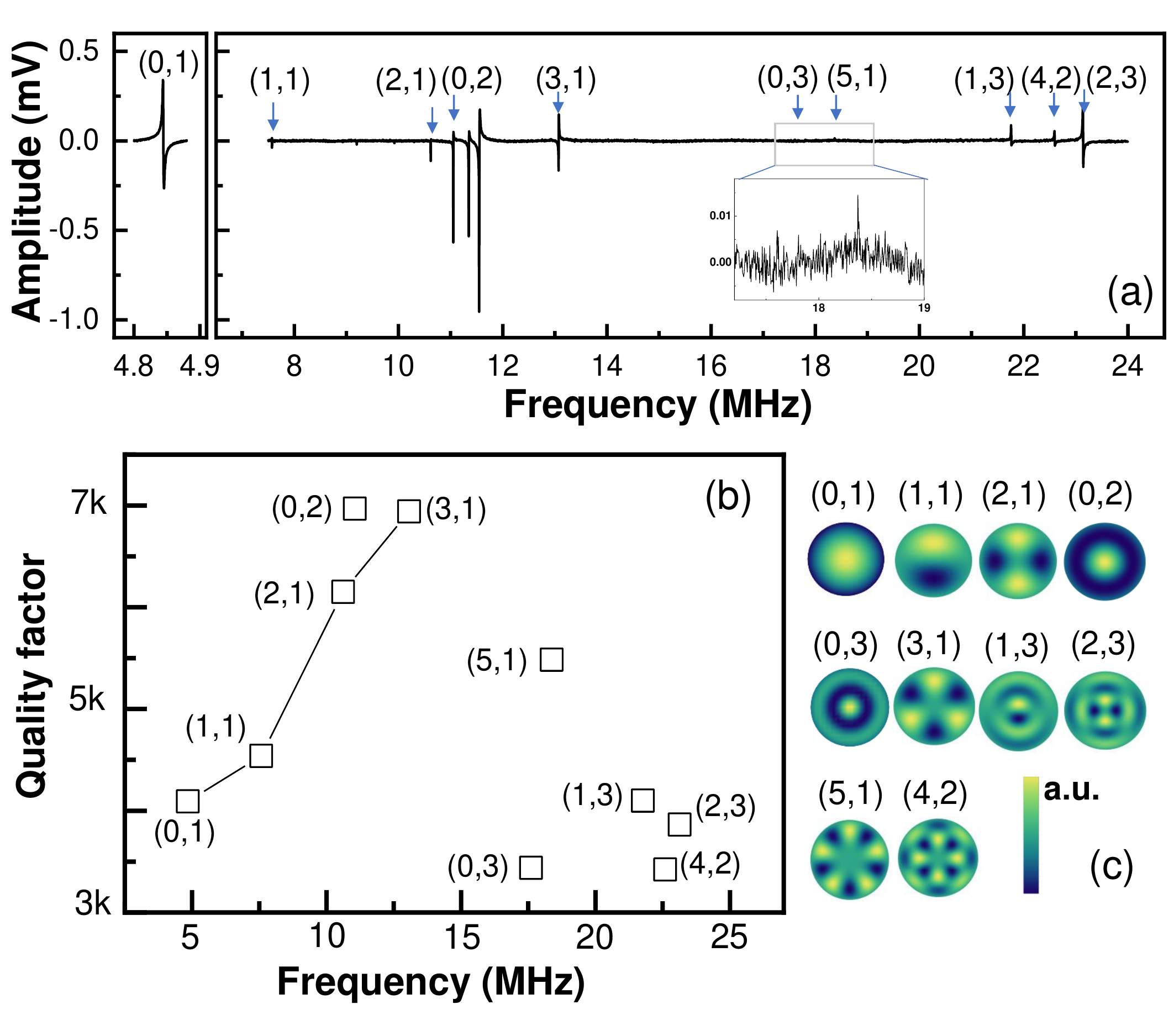}
  \caption{(a) Response of fundamental mode and other higher orders mechanical modes. The device is driven by $V_{ac}$=75 mV and $V_{dc}$=0 V at 4.2 K and measured with microwave reflectometry. The identified eigenmodes, corresponding to the azimuthal and radial mode number, have been marked. There are several unknown modes,  requiring further investigations in the future. (b) Quality factor value corresponds to each mode, which is measured under weak ac signals to avoid the nonlinearity. (c) Mechanical displacements correspond to different modes, depicted here from simulations}
  \label{sch:mutimodes}
\end{figure}

We characterize eigenfrequencies of a low stress SiN drum with a diameter D $\approx$ 40 $\mu$m at 4 K temperature by using the microwave reflectrometry scheme, in which the mechanical modes are excited one after the other by sweeping up the frequency of a large ac signal. Ten eigenmodes have been identified from 4.8 MHz to 24 MHz by comparing the value of these measured harmonics over the fundamental mode with the value of $ \zeta_{nm}/\zeta_{01}$. Thanks to the large coupling capacitance between the drum and its gate and suitable mechanical properties of the drum, we could detect such tiny differences in the displacement generated by mechanical modes, as shown in Fig. \ref{sch:mutimodes}. (c). We examine the $Q$ of each mode in the linear region, as shown in Fig. \ref{sch:mutimodes}. (b), and find the $Q$ of low azimuthal harmonics corresponding to $m=1$ rising as their azimuthal number $n$ increases. A similar phenomenon has been observed in SiN drums with large diameters and is attributed to the destructive interference of the radiated waves \cite{wilson2011high, adiga2011modal}. For higher harmonics, e.g. $n=2, m=3$, the $Q$ value falls back to a value similar to the fundamental mode. It could be due to the mechanical displacement induced local strain, which has been well studied in doubly-clamped SiN beams \cite{unterreithmeier2010damping}. The $fQ$ factor of this low stress SiN drum reaches $\sim$10$^{11}$ Hz at 4 kelvin, which is of the same order as for the high stress SiN drum at RT. 

\begin{figure}
  \includegraphics[width=0.55\textwidth]{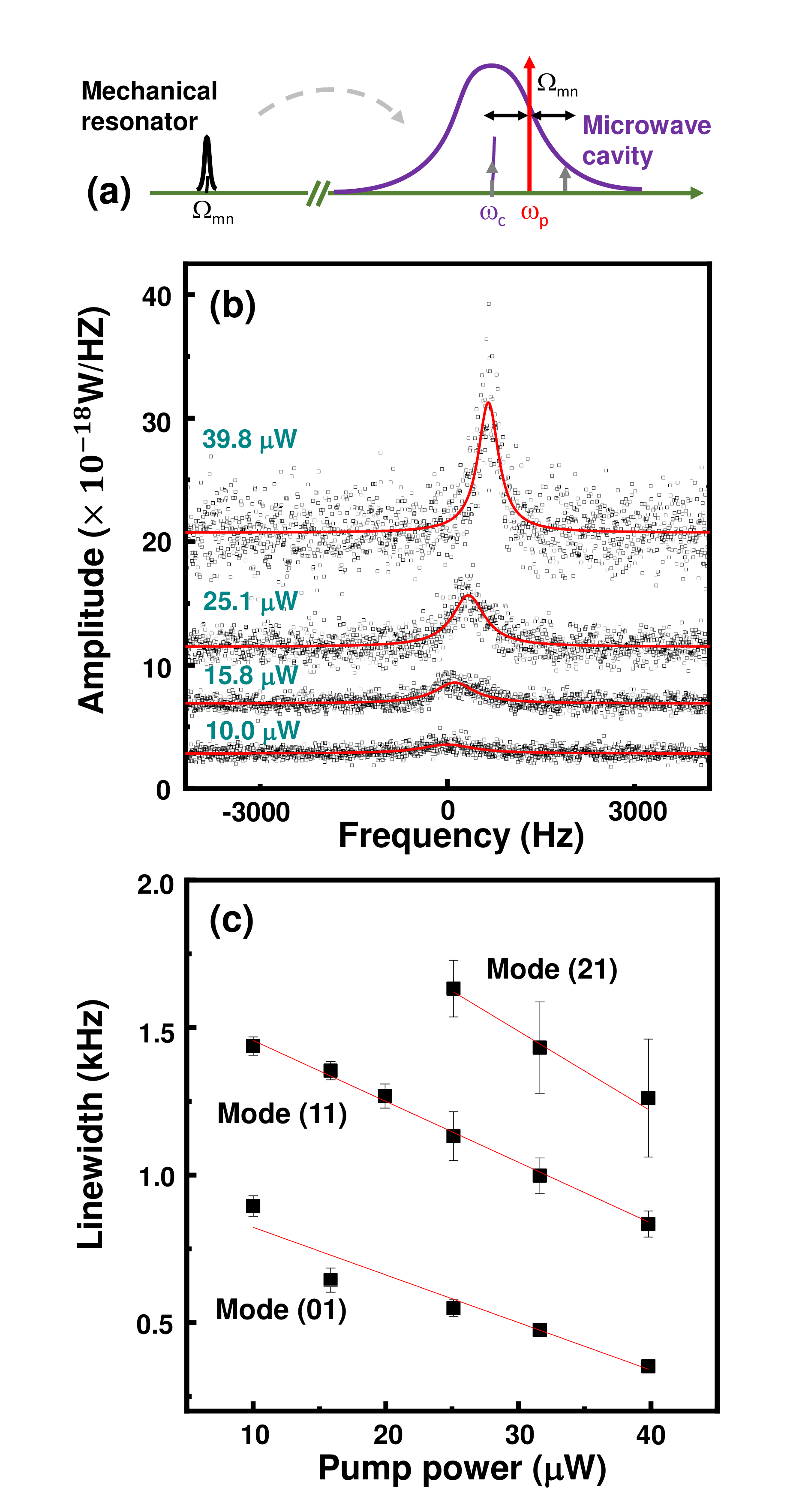}
  \caption{(a) Diagram of blue sideband pumping. The damping rate of the microwave cavity $\kappa_c \approx$ 61 MHz, yielding unresolved sideband condition \citep{zhou2021electric}.  (b) Brownian  spectrum of the first mode (01) measured on the Stokes peaks, corresponding to different pump powers. (c) Effective linewidth of different mechanical modes as function of the pump power. All the optomechanical measurement is performed at the $\approx$ 4 K  temperature.}
  \label{sch:spectrum}
\end{figure}
%
%%%%
%%%%%%%%%%%%%%%%%
Besides microwave reflectometry, our measurement design also gives the ability to interact with the device through  an optomechanical platform thanks to the capacitive coupling to a microwave cavity, as shown in Fig.\ref{sch:setup}.(d). It offers a method to project thermal Brownian motion of the mechanical resonator onto Stokes and anti-Stokes sideband peaks detected in the microwave spectrum, by means of an appropriate pumping of the cavity \citep{aspelmeyer2014cavity}. In addition, when the cavity is sideband pumped at $\omega_c \pm \Omega_m$, optical damping modifies the intrinsic mechanical linewidth$\gamma_m$ by an amount $\delta\gamma_m$, yielding the effective linewidth $\gamma_{eff}$=$\gamma_m$+$\delta\gamma_m$. Under blue sideband pumping (i.e. at $\omega_c+\Omega_m$), the $\delta \gamma_m$ is given by
\begin{equation}
  \delta\gamma_m=-G^2 (\frac{\hbar}{2 m_{eff}\Omega_{nm}})(\frac{4P_{in}}{\hbar\omega_c ({\Omega_{nm}}^2+(\frac{\kappa_c}{2})^2)})(1-\frac{1}{1+(\frac{4\Omega_{nm}}{\kappa_c})^2}), \label{eqn:opticalSpring}
\end{equation}
where $G$=-$\partial\omega_c / \partial x$ is the coupling strength between the mechanical resonator and the cavity, $\hbar$ is the Planck constant, and $P_{in}$ is the input pump power \citep{zhou2021electric}. Obviously, the $\gamma_{eff}$ decreases with increasing pumping power  $P_{in}$, since the $\delta\gamma_m$ is negative. In order to demonstrate this phenomenon, we pumped the cavity at its blue sideband and measured the Stokes peak around $\omega_c$, as shown in Fig.\ref{sch:spectrum} (a). Figure \ref{sch:spectrum}. (b) shows spectra density of the first mechanical mode $\Omega_{01}$ corresponding to different pump powers. The optical damping effect is clearly observed. %We would like to reminder readers that it is not under the sideband resolved condition $\kappa_c \ll \Omega_{nm}$, because the cavity have a large damping rate $\kappa_c \approx 6.1 \times 10^7$.

\begin{figure}
  \includegraphics[width=0.55\textwidth]{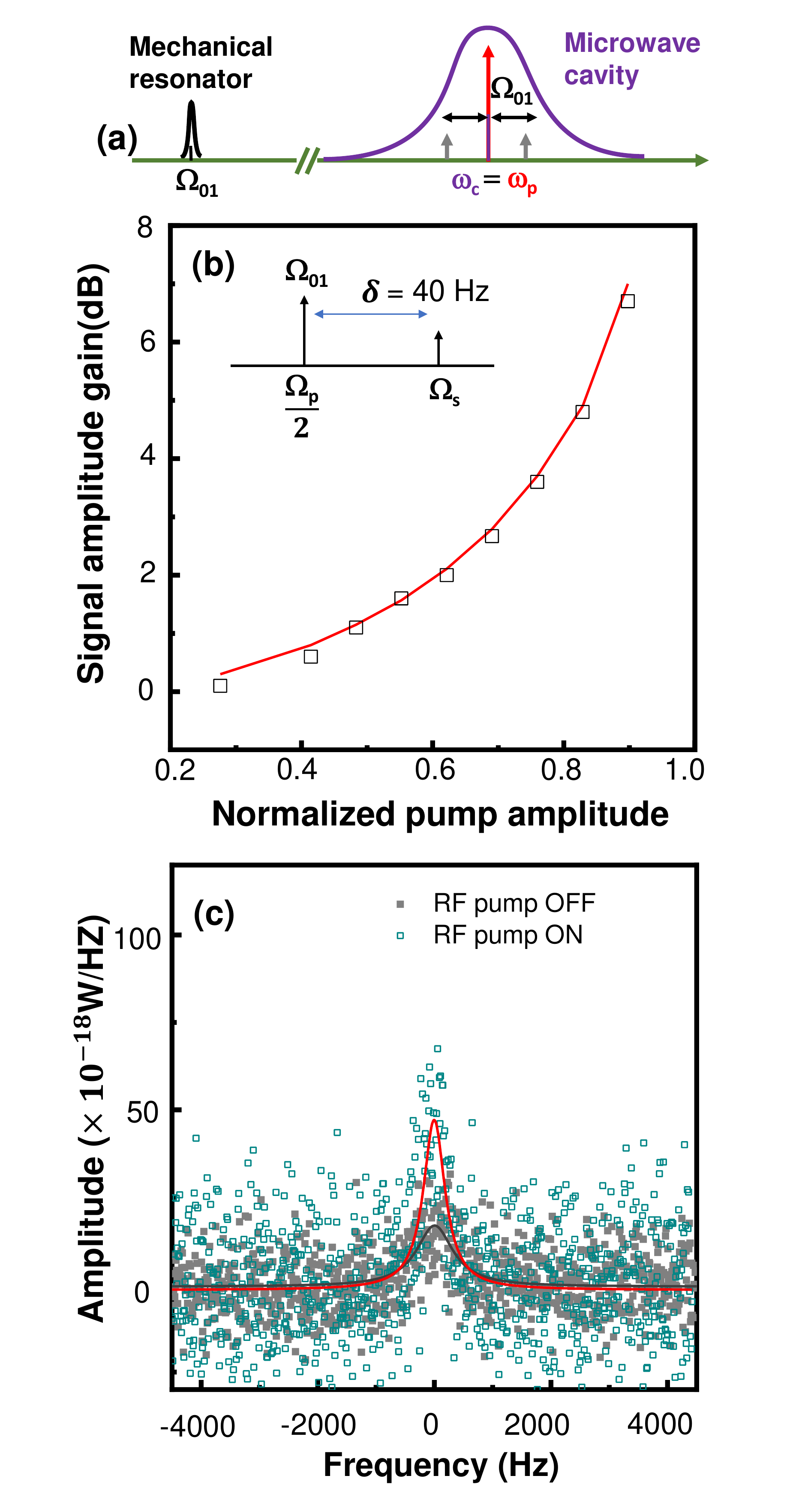}
  \caption{(a) Schematic diagram of microwave optomechanical operations. (b) Signal gain as function of the normalized pump amplitude, measured at non-degenerate condition. The mechanical resonator is parametrically pumped at two times the resonance frequency, $\Omega_{p}$ = $2\Omega_{01}$ and is probed by a small ac signal with $\delta$ = 40 Hz detuning from $\Omega_{p}/2$. Black squares are the measurement results of the signal gain corresponding to the ratio between pump on and off. The red line is theoretical fitting curves (see SI: Theory of non-degenerate mechanical parametric amplification) (c) Power spectral density of mechanical resonator measured at anti-Stokes peak $\omega_p+\Omega_{01}$ when cavity was pumped at its resonance frequency $\omega_c$. Gray squares are spectral density of the thermal noise occupied by the mechanical resonator, without parametric pump. Dark blue squares are measured when the mechanical resonator is parametrically pumped at $2\Omega_{01}$ with a normalized amplitude of the $\sim$ 0.92.}
  \label{sch:rfpump}
\end{figure}
The coupling strength $G$ is an important parameter in optomechanical systems, which can be obtained from the slopes of the $\gamma_{eff}$ versus $P_{in}$ curves based on the Eq.\ref{eqn:opticalSpring}. Figure \ref{sch:spectrum}.(c) shows the linewidth of three different mechanical modes under blue sideband pumping condition. Through fitting these experimental results, slopes of -16.1, -20.6, -27.1 kHz/$\mu$W are obtained corresponding to the mode (01), (11) and (21). The $m_{eff}$ in Eq.\ref{eqn:opticalSpring}, corresponding to each mechanical mode, has been calculated by taking a mode-shape factor 0.2695, 0.2396 and 0.2437 to multiply the drum's mass \citep{hauer2013general}. These values give the coupling strength $G\approx$ 645 kHz/nm, $\approx$ 591 kHz/nm and $\approx$ 647 kHz/nm for the mode (01), (11) and (21). From the definition of $G$, we can re-write it as $G$ = $C_{g}{\omega_c}^2Z / (4d)$, where $Z$ is the impedance of microwave cavity and $d$ is the distance between SiN drum and its top-gate. Following this expression, these values of the coupling strength $G$ yield the $d$ in the range from 450 nm to 500 nm, which is in accordance with our nanofabrication parameter. The coupling strength of our drum is $\sim$ 10 times larger than that of a typical doubly clamped beam structure with a similar resonance frequency that is fabricated without the presented demanding fabrication techniques \cite{zhou2019chip, WeigCavityRT}. Although it is two orders smaller than the largest value achieved by Al drumhead resonators (with $d \sim$ 50 nm and $Z \gg$ 50 Ohm) \citep{teufelGround}, we can further increase $G$ by optimizing device design as  $d \sim$ 550 nm and $Z$ = 50 Ohm are far from our technical limit. 
%

%Both ac signals of the pump and the probe are combined with a dc voltage, $V_{dc}$=6V. 
In the final part of this paper, we investigate mechanical parametric amplification performed with the $V_{ac}$ signals modulating the spring constant, while measuring the motion with the
 optomechanical platform. To do so, we gently pump the microwave cavity at the $\omega_c$ to avoid cavity damping effects on the mechanical resonator and measure the anti-Stokes peak around $\omega_c + \Omega_{01}$. First, we perform a non-degenerate mechanical parametric amplification by pumping at $\Omega_p$ = 2 $\Omega_{01}$ and slightly probing the resonator at the frequency $\Omega_s = \Omega_p /2+\delta$ with $\delta$ = 40 Hz, avoiding limitations due to phase sensitivity \citep{olkhovets2001non}. Because of the frequency mixing, the probe signal $\Omega_{01}$ is up converted into the microwave cavity and therefore the signal gain can be measured by tracing the frequency at $\Omega_s$ + $\omega_p$. The signal amplitude gain is given by $G_{sig}$ = $\rvert 1/(1+ \delta k^2 / 16 m^2_{eff} \Omega^2_m (\delta+i \gamma_m/2)^2 ) \rvert $. More derivation details are given in the SI. Figure \ref{sch:rfpump}.(b) shows the signal amplitude gain as function of the pump amplitude, which is fitted by this analytical expression. The non-degenerate amplitude gain is saturated at the 6.7 dB when device is pumped with a normalized amplitude $\sim$ 0.89. This is because the resonance frequency is shifted due to the nonlinearity. Then, when removing this probe signal, the mechanical parametric amplification of the thermal noise itself can also be performed, as shown in Fig. \ref{sch:rfpump}.(c). In conclusion, we have demonstrated the capabilities of the gate coupling scheme in a new type of high-quality SiN mechanical resonators, which support rather good optomechanical coupling and allow to perform mechanical parametric amplification in this hybrid platform.  

%In the future endeavor, 
\section{SUMMARY}
In summary, we reported on a SiN drum electromechanical resonator, which is realized by semiconductor industry compatible nanofabrication process. It leaves desirable space for engineering trade-off among the quality factor, the resonance frequency,   the coupling strength and the nonlinearity. In addition, the circular geometry gives the drum resonator a reasonable nonlinearity and an abundance of mechanical modes, which is attractive for exploring reservoir computing and phononic mode coupling  \citep{dion2018reservoir, mahboob2013multi, sun2016correlated}. More importantly, this device facilitates the applications of SiN based membrane NEMS to be no longer limited to optical fields, but also can be extended to electronics with full electrical integration, such as microwave reflectometry and optomechanical devices. It brings efficiency in sensing and manipulations because of the large coupling between mechanical displacement and electrical signals.

\begin{acknowledgement}

The authors thank fruitful discussions with Eva Weig, Bernard Legrand and Didier Theron.

X.Z. conceived the design of the experiment. X.Z. developed the device with help of M.Z. and S.V. The measurement, data analysis and analytical calculations are performed by X.Z., R.Z. and H.X.. A.F. provided cryostat for the measurement at 4 K temperature. All authors contributed to the interpretation of the results and the manuscript writing. Both X.Z. and E.C. supervised the project. 

We would like to acknowledge support from the STaRS-MOC project No. 181386 from Region Hauts-de-France (X.Z.), the project No. 201050 from ISITE-MOST (X.Z.), the ERC CoG grant ULT-NEMS No. 647917 (E.C.), and the ERC StG grant UNIGLASS No. 714692 (A.F.). The research leading to these results has received funding from the European Union's Horizon 2020 Research and Innovation Programme, under grant agreement No. 824109, the European Microkelvin Platform (EMP). This work was partly supported by the French Renatech network. 

\end{acknowledgement}


\begin{suppinfo}

\begin{itemize}
  \item Filename: Nanofabrication
  \item Filename: The setup and modeling of microwave reflectometry
  \item Filename: Electrostatic force on frequency modulation
  \item Filename: Nonlinear fit Duffing parameter
  \item Filename: Theory of non-degenerate mechanical parametric amplification
\end{itemize}
%

%\DeclareCaptionLabelFormat{myformat}{#1~S#2}
%\captionsetup{labelformat=myformat}

%
\end{suppinfo}


\bibliography{achemso-demo}

\begin{figure}
  \includegraphics[width=8.25cm, height=4.45cm]{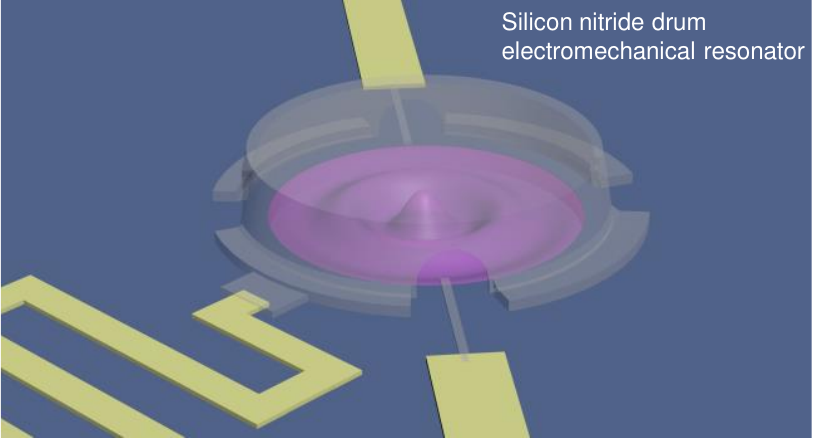}
  \label{sch:TOC Graphic}
\end{figure}

\end{document}

\section{Nanofabrication}
\begin{figure}
  \includegraphics[width=0.95\textwidth]{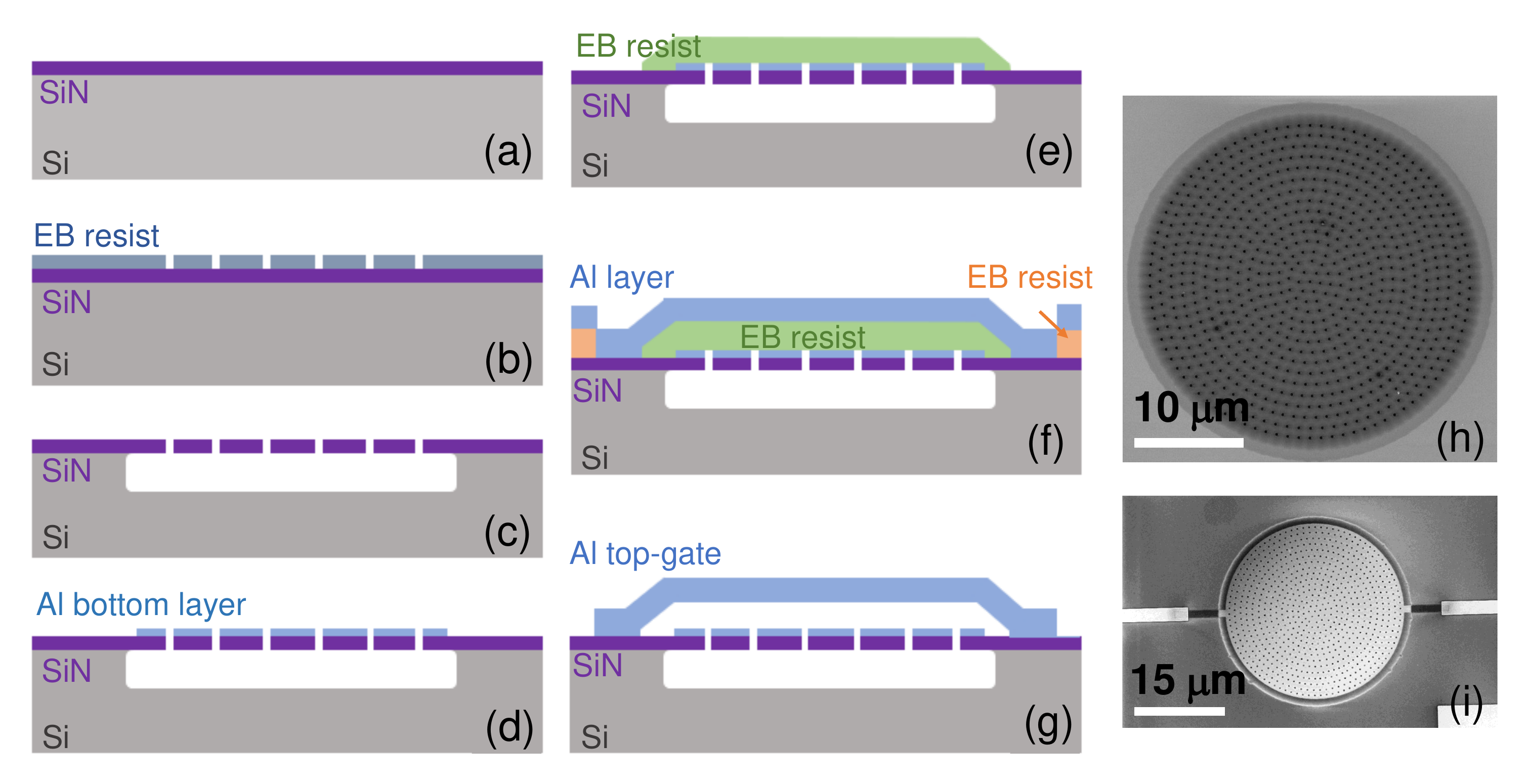}
  \caption{(a-g) Schema of fabrication processes. (a) starting from SiN/Si wafer, (b) pattern holes for etching process (c) release SiN from Si substrate, (d) deposit Al thin film (e) pattern EB resist as a support (f) pattern top-gate and deposited Al metal (g) final device  structure. (h) SEM image of the suspended SiN drum, (i) SEM image of the bottom Al/SiN drum}
  \label{sch:nanofab}
\end{figure}
Figure \ref{sch:nanofab} (a)-(g) shows fabrication process of SiN drum resonators. We define the diameter of the drum by using electron beam (EB) resist CSAR62 to pattern circularly symmetric holes. The drum is released from the Si substrate by reactive ion etching (RIE) of the SiN layer (SF$_6$ : Ar = 10 sccm : 10 sccm, for 6.5 min) through these opened holes, followed by a selective XeF$_2$ silicon etching. These holes occupy about 40 $\% \sim$ 45 $\%$ of the SiN drum area. Its suspended top-gate is fabricated by using EB resist PMMA as a top-gate support through soft-bake at the temperature of 140 $^0$C and reflowed at 180 $^0$C. Then, we deposit the second layer EB resist MMA (methyl methacrylate) on the top of the support resist and pattern the gate structure. For the metal depositions, we first perform Ar ion etching process to clean the sample and then use with electron beam evaporation to deposition the thin films. All bonding pads on the chip are designed to be 50 Ohm impedance for microwave signals.

\section{The setup and modeling of microwave reflectometry}
\begin{figure}
  \includegraphics[width=0.6\textwidth]{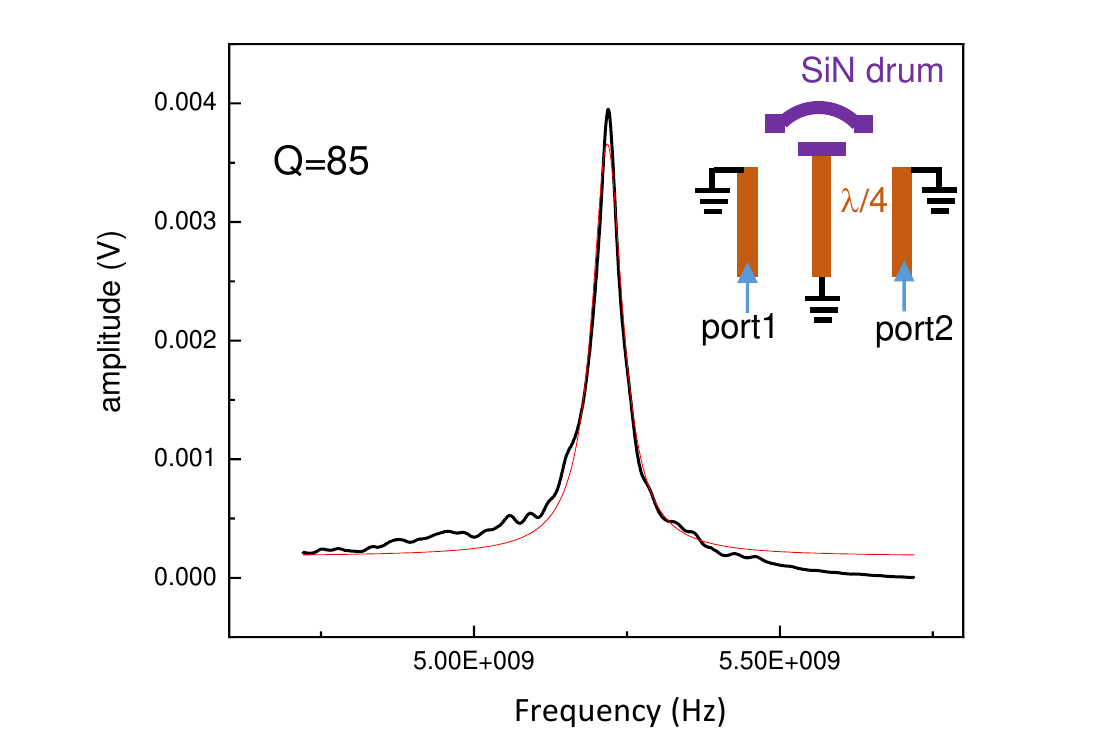}
  \caption{Transmission measurement of the microwave cavity. The inset figure is schema of the cavity, fabricated on the PCB. }
  \label{sch:cavity}
\end{figure}
The microwave cavity is built based on a microstrip line with $\lambda$/4 wavelength on PCB, capacitively coupled to two parallel microstrip lines, as shown in Fig.\ref{sch:cavity}. Its quality factor is obtained through standard transmission measurement. The optomechanical platform is built by connecting the cavity to the top gate of the mechanical resonator through a bonding wire. Both the microwave reflectometry detection line and the driving line connect to the SiN drum through microstrip lines with 50 Ohm impedance on the PCB. The measurement is performed using an ROHDE $\&$ ACHWARZ SMA100B microwave signal generator, an Agilent spectrum analyzer MXA N9020A and a Zurich Instruments UHFLI lock-in amplifier. 

\begin{figure}
  \includegraphics[width=0.6\textwidth]{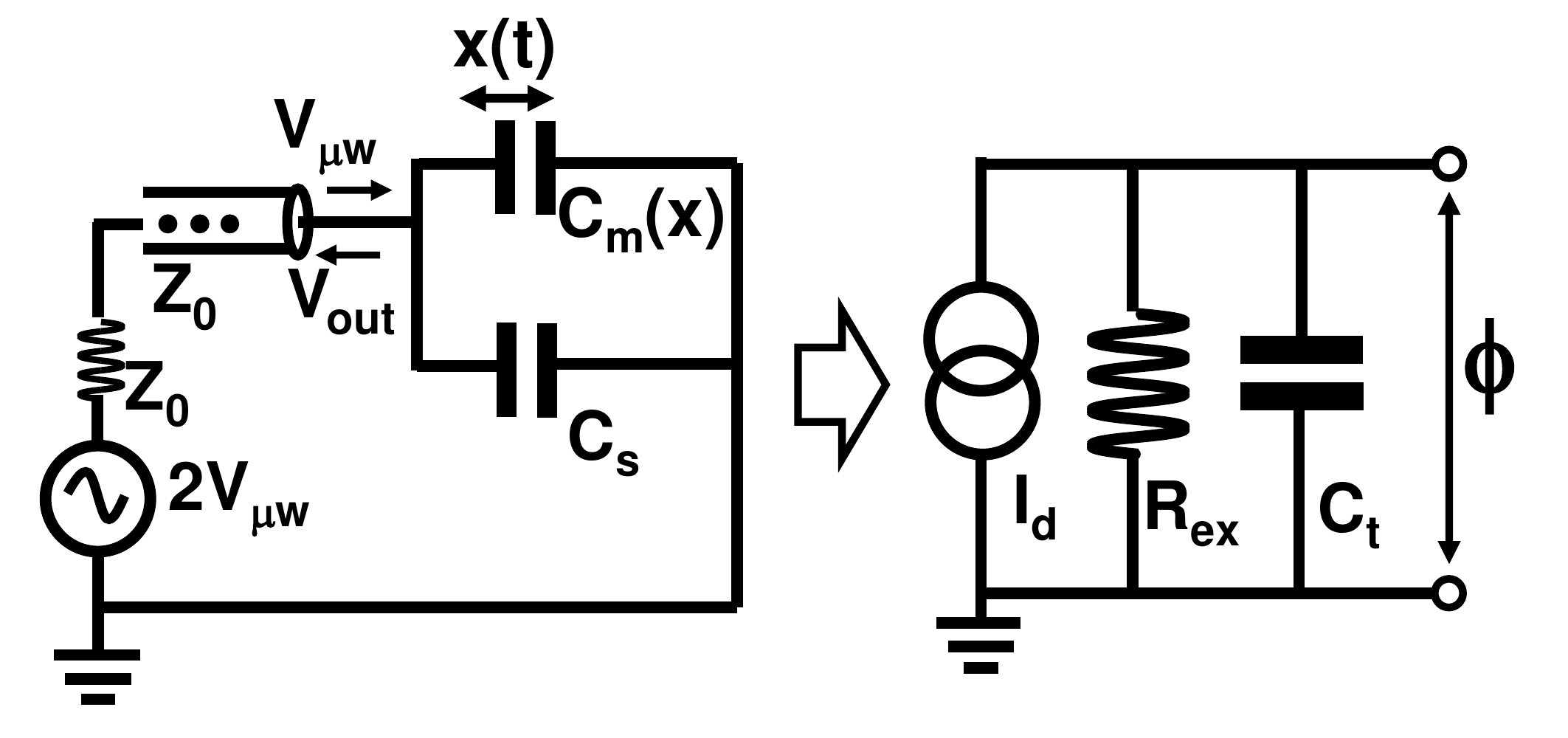}
  \caption{Left, equivalent circuit of reflectometry. The $Z_0$ = 50 Ohm is the impedance of the detection part and $V_{\mu w}$ is the microwave amplitude with frequency $\omega$ delivered to the mechanical resonator. Right, its equivalent parallel $\textit{RC}$ circuit.}
  \label{sch:reflec}
\end{figure}

The mechanical resonator can be treated as a movable capacitor $C_m(x)$, which connects a stray capacitor $C_s$ induced by electrode pads. From Norton’s theorem, it can be transferred to a parallel circuit, as shown in Fig.\ref{sch:reflec}, where the total capacitance $C_t$ = $C_s$+$C_m(x)$ and equivalent resistance $R_{ex}$ = 1/($\omega^2 C^2_t Z_0$). As the $C_m(x)$ is a function of the mechanical displacement $x(t)$, we rewrite it as $C_t$ = $C_{t0}$-$C_m x(t)$/$d$ where $C_{t0}$ is the total static capacitance and $d$ is the initial distance between the drum resonator and its coupling gate. It can be described by a motion equation of the flux $\phi$ biased on the $C_m(x)$,
\begin{equation}
  (C_{t0}-\frac{C_mx}{d})\frac{\partial^2 \phi}{\partial t^2}-\frac{C_m}{d}\frac{\partial x}{\partial t} \frac{\partial \phi}{\partial t}+\frac{1}{R_x} \frac{\partial \phi}{\partial t} = I_d, \label{eqn:refMotion}
\end{equation} 
where $I_d$ = $i \omega C_t V_{\mu w}$, which can be written in form of $I_{d}(t)$ = $\frac{1}{2} I e^{-i\omega t}+ \text{c.c.}$. Here, the $I$ is the complex amplitude of the current. The mechanical displacement is written as $x(t)$ = $\frac{1}{2} A(t)   e^{-i \Omega_m t}+ \text{c.c.}$ with $A$ the (complex) amplitude translated in frequency around the mechanical resonance frequency $\Omega_m$. We can thus find an exact solution using the {\it ansatz}:
\begin{equation}
\label{eqn:ansatz}
\phi (t) = \!\!\! \sum_{n=-\infty}^{+\infty} \!\!\! \phi_n (t)= \!\!\! \sum_{n=-\infty}^{+\infty} \!\! \frac{1}{2}\mu_n (t) e^{-i \left(\omega+ n \Omega_m\right) t}+ \text{c.c.}
\end{equation}
We look for the solution $V_{out}$ corresponding to the frequency $\omega + \Omega_m$, which was measured in the experiment. By solving motion equation \ref{eqn:refMotion} and considering input-output theory \citep{zhou2021electric}, the output signal is given by
\begin{equation}
\label{eqn:solution}
V_{out}=\omega Z_0 \frac{C_m}{d} \frac{\rvert x \rvert}{2} V_{\mu w}.
\end{equation}
Here, we take several reasonable approximations, $\omega$/($\omega + \Omega_m$) $\sim$ 1 and $\omega C_{t0} Z_0 \ll $ 1. This is because the $\omega \gg \Omega_m$ and typical stray capacitance of $C_s$ is in pF range. Based on Eq.\ref{eqn:solution}, we can obtain the mechanical displacement of $x$ through this microwave reflectometry scheme. In the setup, we use the ac signal generated by the Lockin to excite the mechanical displacement $x(t)$ and probe it through a microwave signal $V_{\mu w}$ with power of 10 dBm and  frequency $\omega$ = 4.8 GHz. As explained above, it generates a reflected signal around $V_{out}$. For the measurement, we convert the $V_{out}$ to a low frequency signal around $\Omega_m$ by demodulating it with a signal $\omega$ through a frequency mixer. 
\section{Electrostatic force on frequency modulation}
The electrical signals driving on the mechanical resonator, $V$ = $V_{dc}$ + ${V_{ac}}$, generates the static force $F(x)$ = -$\frac{\partial}{\partial {x}}({C_m(x)V^2}/2)$. It shifts the mechanical resonance frequency through modulating the spring constant of the resonator, 
\begin{equation}
  \Omega_m\approx\Omega_0 (1+\frac{\delta k_{dc}}{2 k}), \label{eqn:shiftOmega}
\end{equation}
where, both $\Omega_0$ and $k$ correspond to the intrinsic resonance frequency and the intrinsic spring constant. The modulation of the spring constant can be obtained by $\delta k_{dc} \approx - \partial F / \partial x$. Then, we treat the resonator as a movable capacitor, which can be depicted as $C_m(x)=C_{m0}/(1+x/d)$.The  $d$ is the initial distance between the drum and its top-gate which determines the initial capacitance $C_{m0}$ and $x$ is the small mechanical displacement. It gives $\partial F / \partial x \approx V^2C_{m0} / d^2 $. 
Thus, Eq.\ref{eqn:shiftOmega} becomes
 \begin{equation}
  \Omega_m\approx\Omega_{0}(1-\frac{C_{m0}V^2}{4 k d^2}). \label{eqn:final}
\end{equation}
If we consider the experimental parameters, $V_{dc} \gg V_{ac}$, and expand $V^2$=$V^2_{dc}+2V_{dc}V_{ac}+V^2_{ac}$, it is not difficult to find that the modulation of resonance frequency is mainly dominated by the $V^2_{dc}$ and therefore the $V^2$ in Eq. \ref{eqn:final} can be replaced by $V_{dc}$. 
\section{Nonlinear fit Duffing parameter}
We consider a Duffing nonlinear mechanical resonator, driven by an external sinusoidal force. It is described by the motion equation
\begin{equation}
  \ddot{x}+\gamma_{m}\dot{x}+\frac{k}{m_{eff}}x+\alpha x^3=\frac{F cos(\omega t)}{m_{eff}}. \label{eqn:nonlinear}
\end{equation}
A well-known secular perturbation theory is used to simplify it to be
\begin{equation}
  \rvert a \rvert^2 = \frac{g^2}{(2 \Omega_x \pm \frac{3}{4} \rvert a \rvert^2)^2 +1},\label{eqn:nonlinearS}
\end{equation}
where, $g$ = $F \sqrt{\alpha Q^3_m / m^3_{eff}} / \Omega^3_m$, $\Omega_x$ = $Q_m (\Omega - \Omega_m) / \Omega_m$, and $a$ = $x \sqrt{\alpha Q_m/ (m_{eff} \Omega^2_m)}$ \citep{lifshitz2008nonlinear}. The $\Omega_m$ is the linear resonance frequency of the resonator. By taking $2 \Omega_x \pm \frac{3}{4}$ = 0, the maximum frequency of the Duffing curve can be obtained. The symbol of $'+'$ and $'-'$ in Eq.\ref{eqn:nonlinearS} corresponds to softening and hardening nonlinear response, respectively. The solutions of the steady-state amplitude for Duffing equation are used for fitting the measurement data, which are obtained by taking the positive real roots of $x$ in solving Eq.\ref{eqn:nonlinearS}. 
\begin{figure}
  \includegraphics[width=0.6\textwidth]{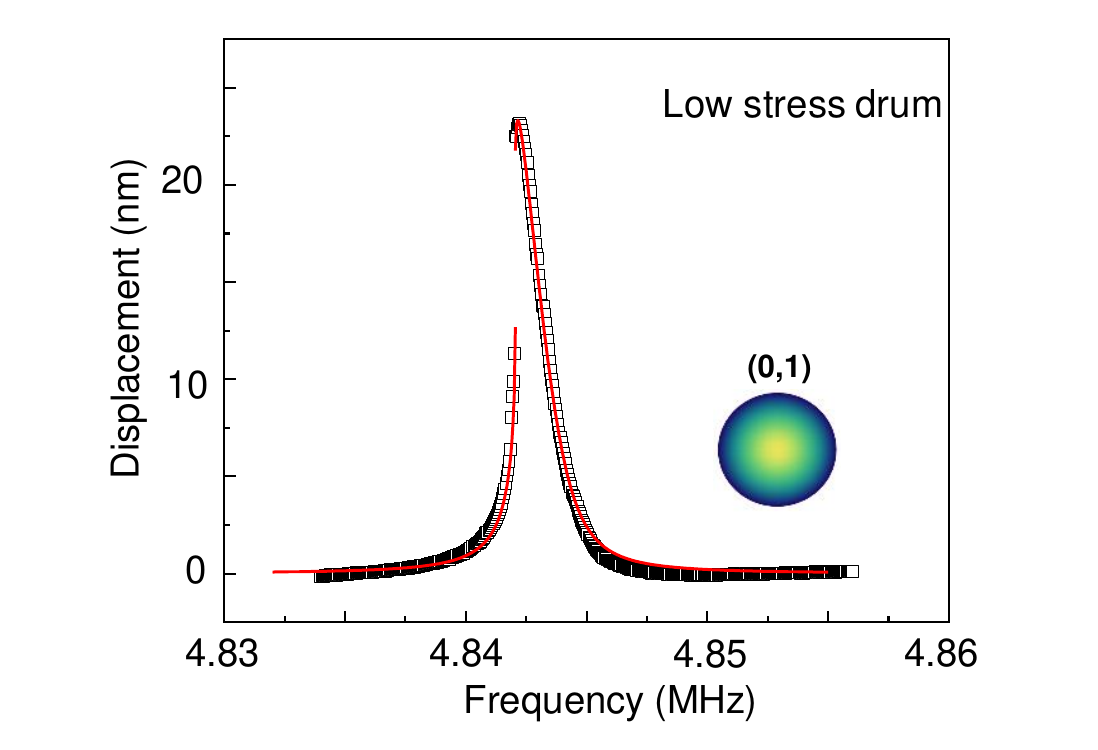}
  \caption{The nonlinear response of the low-stress drum driven under 220 pN force. The red line is the fitting curve.}
  \label{sch:nonliear}
\end{figure}

By using Eq.\ref{eqn:solution}, we convert the amplitudes of the electric signals $V_{out}$ to the mechanical displacement $x$. The values of $C_m$ and $d$ are calculated based on nanofabrication parameters. Figure \ref{sch:nonliear} shows a nonlinear response of the low-stress drum driven by an electrical force $\approx$ 220 pN ($V_{dc}$ = 0.7 V, $V_{ac}$ = 10 mV). The fitting results give a duffing nonlinear parameter $\alpha \approx$ 2.2 $\times$ 10$^{27} m^{-2} s^{-2}$, the fitted quality factor $Q_{fit} \approx$ 4000 and the fitted mechanical resonance frequency $\Omega_{fit}\approx$  4.843 MHz. In all nonlinear parameter fits, both $Q_{fit}$ and $\Omega_{fit}$ are within the $5 \%$ discrepancy with the measured values. 
\section{Theory of non-degenerate mechanical parametric amplification}
We start with the well-known motion equation of a parametrically pumped mechanical resonator, 
\begin{equation}
  m_{eff} \ddot{x}+m_{eff}\gamma_m\dot{x}+(k+\delta k Sin(\Omega_p t))x = F_0 Cos(\Omega_s t), \label{eqn:degM}
\end{equation}
where $\delta k$ is the variations of the spring constant modulated by a pump with frequency $\Omega_p$ and  $F_0 Cos(\Omega_s t)$ is the driving force. We consider following frequency detuning condition: $\Omega_m$ - $\Omega_p /2$ = $\Delta$, $\Omega_s$ - $\Omega_p /2$ = $\Omega_p /2$ - $\Omega_i$ = $\delta$, where $\Omega_m$ is the resonance frequency of the interested mechanical mode. Because the frequency detuning, an idler signal is generated at $\Omega_i$. Thus, we look for the solutions of the $x$ corresponding to the frequency of the driving signal $x_s$=$\mu_s e^{-i \Omega_s t}/2 + c.c$ and to the idler $x_i$=$\mu_i e^{-i \Omega_i t}/2 + c.c$. Then, Eq.\ref{eqn:degM} becomes
\begin{equation}
  2 m_{eff} \Omega_m \chi_s \mu_s - \frac{\delta k}{2 i} \mu^*_i = F_0, \label{eqn:md}
\end{equation}
\begin{equation}
  2 m_{eff} \Omega_m \chi_i \mu_i - \frac{\delta k}{2 i} \mu^*_d= 0, \label{eqn:mi}
\end{equation}
where, 
\begin{equation}
  \chi_s = \Delta - \delta - i \frac{\gamma_m}{2}, \label{eqn:chd}
\end{equation}
\begin{equation}
  \chi_i = \Delta + \delta - i \frac{\gamma_m}{2}. \label{eqn:chi}
\end{equation}
The amplitude gain $G_{sig}$ is given by the ratio of the amplitude $\rvert \mu_s \rvert$ with and without the parametric modulation $\delta k$,
\begin{equation}
G_{sig} =\rvert  \frac{1}{1-\frac{\delta k^2}{16 m^2_{eff} \Omega^2_m \chi_s \chi^*_i }}\rvert. \label{eqn:gain}
\end{equation}
The critical driving amplitude, which starts to drive the resonator into self-auto oscillation state, is obtained by taking the denominator of Eq.\ref{eqn:gain} = 0 at the degenerate condition $\delta$ = 0. It gives the critical modulation : %rewrite !!! phase 
\begin{equation}
\delta k_{c} = \frac{2k}{Q}\sqrt{1+\frac{4 \Delta^2}{\gamma^2_m}}. \label{eqn:deltaK}
\end{equation}
The pump amplitude is normalized by the critical pump amplitude $p_c$ triggering the self-auto oscillation, which is obtained from the expression of the $\delta k \approx (V_{pump} + V_{dc})^2 C_{m0} / d^2$. In order to generate non-degenerate parametric amplification in the experiment, we probe the drum resonator by $V_{probe} Cos(\Omega_{s} t)$ with a frequency detuning $\delta$=40 Hz from the $\Omega_{01}$ and pump the spring constant by $V_{pump} Cos(2 \Omega_{01} t)$. Here, the $\Omega_{01}$ is the resonance frequency of the first mode. An ac signal with amplitude $V_{probe}$ = 200 $\mu V$ is used to probe the drum resonator and the pump amplitudes $V_{pump}$ are in the range from 5 mV to 18 mV, mixed with $V_{dc}$ = 6 V. The measurement results are fitted by taking $\Delta$ = 0 and $k \sim$ 21 N/m. 

\bibliography{supplement}